\documentclass[conference,letterpaper]{IEEEtran}

\usepackage[utf8]{inputenc} 
\usepackage[T1]{fontenc}
\usepackage{url}
\usepackage{ifthen}
\usepackage{cite}
\usepackage[cmex10]{amsmath} 

\usepackage{cite}
\usepackage{mwe}
\usepackage{subfig}
\usepackage{setspace}
\usepackage{algorithm}
\usepackage{algorithmic}
\usepackage{float}%
\usepackage{commath}
\usepackage{ifpdf}
\usepackage{tabularx}
\usepackage{cuted}
\usepackage{lipsum}
\usepackage{diagbox}
\usepackage{multirow}
\usepackage{xcolor}
\usepackage{romannum}
\usepackage{etoolbox}
\usepackage{mathtools}
\usepackage{amsfonts, amsmath, amssymb}
\usepackage{bbm}
\usepackage{bm}
\usepackage{graphicx, wrapfig, color}
\usepackage{hyperref}
\usepackage[normalem]{ulem}

\usepackage{mathtools}
\mathtoolsset{showonlyrefs}

\usepackage{annotate-equations}
\usepackage[skins,listings]{tcolorbox}

\usepackage{comment}

\allowdisplaybreaks

\usepackage{xcolor}

\usepackage{tikz}
\usetikzlibrary{tikzmark} 
\usepackage{hf-tikz} 

\begin{document}
\title{Interpreting Deepcode, a learned feedback code} 



\author{\IEEEauthorblockN{Y. Zhou, N. Devroye, Gy. Tur\'an$^*$,  and M. \v Zefran }
\IEEEauthorblockA{University of Illinois Chicago, Chicago, IL, USA}
\IEEEauthorblockA{$^*$ HUN-REN-SZTE Research Group on AI}
\IEEEauthorblockA{\{yzhou238, devroye,  gyt, mzefran\}@uic.edu}
}


\maketitle

\begin{abstract}
 Deep learning methods have recently been used to construct non-linear codes for the additive white Gaussian noise (AWGN) channel with feedback. However, there is limited understanding of how these black-box-like codes with many learned parameters use feedback. This study aims to uncover the fundamental principles underlying the first deep-learned feedback code, known as Deepcode, which is based on an RNN architecture. Our interpretable model based on Deepcode is built by analyzing the influence length of inputs and approximating the non-linear dynamics of the original black-box RNN encoder. Numerical experiments demonstrate that our interpretable model -- which includes both an encoder and a decoder --  achieves comparable performance to Deepcode while offering an interpretation of how it employs feedback for error correction. \footnote{This work was supported by NSF under awards 1900911, 2217023, and 2240532, and by the AI National Laboratory Program (RRF-2.3.1-21-2022-00004). The contents of this article are solely the responsibility of the authors and do not necessarily represent the official views of the NSF.}

\end{abstract}

\section{Introduction}
Although it is known that feedback does not increase the capacity of memoryless channels, it can reduce the coding complexity and improve reliability.  Moreover, for finite blocklengths and fixed rates, constructing codes that achieve the smallest bit or block error rate remains open. This has motivated the construction of  deep-learned error-correcting codes (DL-ECC) \cite{kim2018deepcode, safavi2021deep, mashhadi2021drf, shao2023attentioncode, ozfatura2022all, kim2020deepcode, ben2020simple, kim2023robust} in which the encoding and decoding functions are parameterized by a (usually) very large number of parameters in a neural network architecture, which are then learned by adjusting these to numerically minimize a loss function. This approach differs markedly from previous feedback coding schemes \cite{schalkwijk1966coding,shayevitz2011optimal, ben2017interactive, chance2011concatenated, mishra2023linear} which are analytically constructed. 
Due to the increasing model complexity, it is difficult to understand how these codes accomplish error correction, and this leads us to perceive learned models as ``black boxes''.
 Such an  understanding is  important to a) build trust in these models, b) identify their weaknesses, and c) reveal how feedback is used, potentially pointing us in the direction  (similar in spirit to \cite{lian2018can}) of new (possibly non-linear) coding schemes. 
We aim to open the black box and \emph{interpret} a prominent early example of the learned feedback encoders and decoders: Deepcode~\cite{kim2018deepcode}. 

Deepcode is a recurrent neural network (RNN)-based non-linear coding scheme for AWGN channels with passive feedback. Its experimental error performance was shown~\cite{kim2018deepcode} to outperform that of Schalkwijk-Kailath (SK)~\cite{schalkwijk1966coding} and Chance-Love (CL) \cite{chance2011concatenated} schemes in the case of passive noisy feedback.  
More recently, a state propagation-based non-linear feedback code based on RNNs has emerged  \cite{kim2023robust} which encompasses SK, CL, and Deepcode and is especially robust to feedback noise levels;  but it is still unclear exactly how this scheme utilizes the feedback symbols for error correction.

 The relevant notion of interpretation depends on context. Here we aim to provide an understanding of the deep-learned models similar to the analytically constructed feedback coding schemes. Deepcode \cite{kim2018deepcode} was the first to offer a limited interpretation of the underlying functioning of encoders / decoders through scatter-plots and coupling. We significantly expand the understanding of how Deepcode works here.

{Some approaches have been proposed to understand RNNs  \cite{sussillo2013opening,karpathy2015visualizing,choe2017probabilistic }, but none is immediately applicable in this novel DL-ECC setting. {The validity of explanations can be controversial as well \cite{Jain19,Wiegr}.}
Unlike previous classification tasks, we focus on understanding a pair of 
RNNs,
an encoder and a decoder, which are jointly learned in the presence of channel noise,   to transmit and decode exponentially many messages.}
Our prior work \cite{devroye2022interpreting, mulgund2022evaluating, devroye2023interpreting, mulgund2023decomposing} has focused on interpreting TurboAE \cite{jiang2019turbo} -- a deep-learned forward error-correcting code based on convolutional neural networks (CNNs) placed into a Turbo-code-like architecture with an interleaver,  in the absence of feedback.  The coding structure of the feed-forward, CNN-based model with an interleaver varied markedly from the RNN and feedback structure of Deepcode, and while some methods carry over, new techniques must be developed for interpretation in the feedback setting. 
 
 {\bf Contributions.} We suggest several tools for understanding how Deepcode works, including simplifying RNN through dimension reduction and pruning, outlier analysis, architecture-based insights, influence length, and nonlinear {input-output map} approximations.
 Our interpretation efforts culminate by proposing an interpretable encoder and decoder approximation yielding an understanding\footnote{An explicit scheme with a small number of learnable parameters and an interpretation of how error-correction is performed at the encoder and decoder.} of how feedback is used to correct errors in Deepcode, with competitive, and sometimes even superior BER performance in noisy or noiseless feedback. 

{\bf Notation}: subscripts $i,j$ represent time and bit indices, respectively.
Vectors of random variables are expressed in bold, with superscripts  indicating their lengths. 
 $K$ and $N$ represent the lengths of message bits and codewords, respectively. The coding rate is  $r = K/N$. $\text{SNR}_f$ and $\text{SNR}_{fb}$ represent the forward and feedback channel Signal-to-Noise Ratios, respectively. $\mathbb{R}^n$ represents $n$-dimensional real vectors.  $\mathbb{F}_2$ denotes the finite field with elements 0 or 1 and addition  $\bigoplus$.
 Function $\mathbb{I}(x) = 1$ if the argument is true (non-negative), 0 else. 


\section{System Model}

\begin{figure}
  \centering
  \begin{tikzpicture}
    \node (encoder) at (2,0) [draw, rectangle] {$\text{Encoder}$};
\node (plus1) at (4,0) [draw, circle] {$+$};
\node (decoder) at (6,0) [draw, rectangle] {$\text{Decoder}$};
\node (plus2) at (4,-1.5) [draw, circle] {$+$};

\draw[->] (0,0) -- (encoder) node[midway, above] {$\mathbf{b}$};
\draw[->] (encoder) -- (plus1) node[midway, above] {$x_i$};
\draw[->] (plus1) -- (decoder) node[midway, above] {$y_i$};
\draw[->] (decoder) -- (8,0) node[midway, above] {$\mathbf{\hat{b}}$};
\draw[->] (4,0.7) -- (plus1) node[pos=0.2, above] {$n_i\sim \mathcal{N}(0,\sigma_f^2)$};
\draw (5,0) -- (5,-1.5) node[midway, right] {$y_{i-1}$};
\draw[->] (5,-1.5) -- (plus2);
\draw (plus2) -- (2,-1.5) node[midway, above] {$\tilde{y}_{i-1}$};
\draw[->] (2,-1.5) -- (encoder);
\draw[->] (4,-2.2) -- (plus2) node[pos=0.2, below] {$\tilde{n}_{i-1}\sim \mathcal{N}(0,\sigma_{fb}^2)$};
  \end{tikzpicture}
  \caption{\small AWGN channel with passive noisy feedback.}
  \vspace{-6mm}
  \label{fig:awgn}
\end{figure}
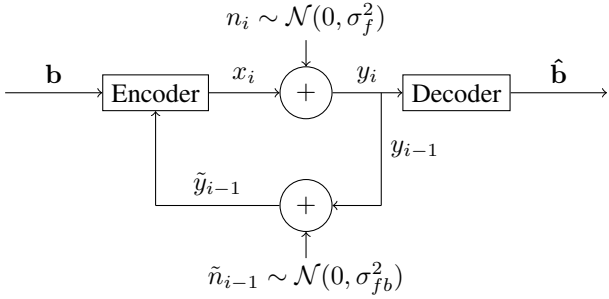



Fig. \ref{fig:awgn} shows the AWGN channel with passive noisy feedback. The message bits $\mathbf{b}\in \mathbb{F}_2^K$ are sent through $N$ time steps. At each time instant $i\in\{1, \ldots, N\}$, the forward channel is characterized by $y_i = x_i + n_i$, where $x_i \in \mathbb{R}$ is the transmitted symbol, and $n_i \sim {\cal N}(0,\sigma_f^2)$ is the Gaussian noise, independent and identically distributed (iid) across time steps. The receiver sends channel outputs back to the transmitter with one unit delay through the noisy channel: $\tilde{y}_{i-1} = y_{i-1} + \tilde{n}_{i-1}$. Here, $\tilde{n}_{i-1}\sim \mathcal{N}(0,\sigma_{fb}^2)$ is also an iid Gaussian noise.

Encoding functions $f_i$ map the message bits and feedback to transmitted codewords (joint coding and modulation), denoted as $x_i = f_i(\mathbf{b}, \mathbf{\tilde{y}}^{i-1})$. The decoding function $g$ maps the channel outputs to estimated message bits $\mathbf{\widehat{b}} = g(\mathbf{y}^N)\in \mathbb{F}_2^K$.
We impose an average power constraint $\frac{1}{N}\mathbb{E}\left[\lVert \mathbf{\mathbf{x}} \rVert_2^2\right]\leq 1$ where $\mathbf{x} = (x_1, \cdots, x_N)$. Performance is measured through the bit error rate $BER=\frac{1}{K}\sum_{i=1}^{K}\mathbb{P}(b_i\neq \widehat{b}_i)$. 

\begin{figure}[ht]
\vspace{-4mm}
    \centering
    \includegraphics[width=\columnwidth]{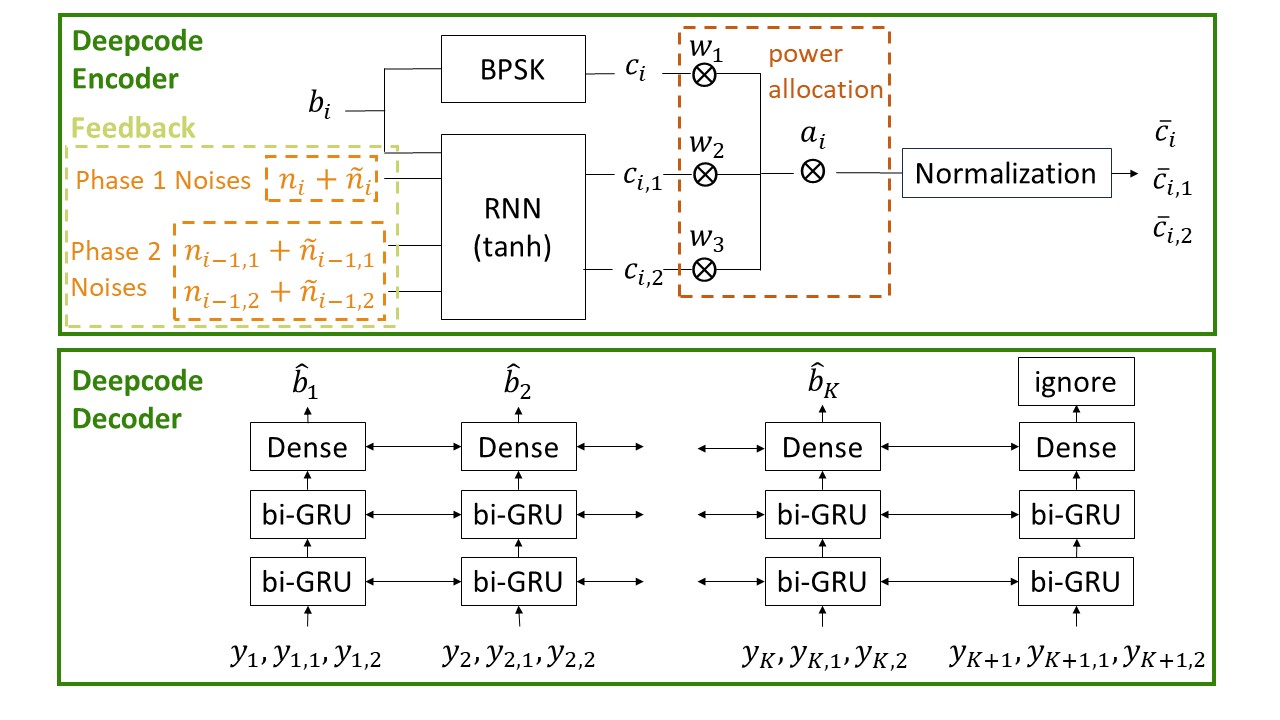}
    \caption{Deepcode encoder (above) and decoder (below). {Here, $i \in \{1, \ldots, K + 1\}$ because the message bits are padded with a zero. When $i = 1$, the initial value for phase 2 noises is $0$.}}
    \vspace{-4mm}
    \label{fig:deepcode_structure}
\end{figure}

Deepcode \cite{kim2018deepcode} is a DL-ECC designed for this AWGN channel with rate $1/3$ (Fig.~\ref{fig:deepcode_structure}). The encoding scheme contains two phases. In the first phase, the $K$ message bits $\mathbf{b}$ under BPSK modulation $\mathbf{c}= 2\mathbf{b}-1 \in \{-1, 1\}^K$ are transmitted, uncoded, through the channel,  yielding the feedback $\mathbf{\tilde{y}} = \left[\tilde{y}_1, \cdots, \tilde{y}_K\right]$. The encoder stores noises in the first phase $\mathbf{n} + \mathbf{\tilde{n}} = \mathbf{\tilde{y}} - \mathbf{c} \in \mathbb{R}^K$ for later use. In the second phase, the encoder uses a directional RNN with a $\tanh$ activation function  and linear combination layer to sequentially generate $2K$ parity bits $c_{i, 1}$ and $c_{i,2}$ where $i\in \{1, \ldots, K\}$. At time instant $i$, we define the input to the RNN as $\mathbf{P}_i = \left[b_i, n_i + \tilde{n}_i, n_{i-1, 1} + \tilde{n}_{i-1, 1}, n_{i-1, 2} + \tilde{n}_{i-1, 2} \right]$ which contains the current message bit ($b_i$), the noise in the first phase ($n_i + \tilde{n}_i$), and the feedback noises resulting from the transmission of parity bits in the second phase ($n_{i-1, j} + \tilde{n}_{i-1, j}$ $ = \tilde{y}_{i-1, j} - c_{i-1, j}$, $j\in \{1, 2\}$).  {The codewords are denoted as $\mathbf{X}^N =$ $ \left[c_1,\cdots,c_K, c_{1,1}, c_{1,2},c_{2,1}, c_{2,2},\cdots, c_{K,1}, c_{K,2} \right] $}.
The decoding scheme uses a two-layered bidirectional gated recurrent unit (GRU) to estimate message bits $\mathbf{\widehat{b}}$ from the noisy codewords $\mathbf{Y}^N$. In what follows, we initially focus on the noiseless feedback case, where $\tilde{n}_i = \tilde{n}_{i-1,1} = \tilde{n}_{i-1,2}=0$. We later extend our scheme to noisy feedback.


In Deepcode, zero padding is applied to the last message bit to reduce the error. Codewords are assigned different learned power weighting parameters $\mathbf{w}$ and $\mathbf{a}$ to balance errors. Finally, the codewords are normalized to satisfy the power constraint (Fig.~\ref{fig:deepcode_structure}). The encoder and decoder are trained jointly to minimize the binary cross-entropy (BCE). \emph{``TensorFlow Deepcode''} will refer to the original Deepcode implementation with $N_h = 50$ hidden states. We implemented our \emph{``Pytorch Deepcode''} based on the TensorFlow Deepcode\footnote{Code is available at \url{https://github.com/zyy-cc/Deepcode-Interpretability} and \url{https://github.com/hyejikim1/Deepcode}.}.




\section{Model Reduction}
{Since Deepcode has $N_{h} = 50$ hidden states, the model becomes relatively complex, with over $65,000$ parameters. This makes direct interpretation challenging. 
In this section, we perform model reduction (through dimension reduction and pruning) to find a model of much smaller dimension / fewer parameters without sacrificing performance. }


\subsection{Dimension Reduction}

\begin{figure}[ht]
    \vspace{-5mm}
    \centering
    \includegraphics[width=0.4\textwidth]{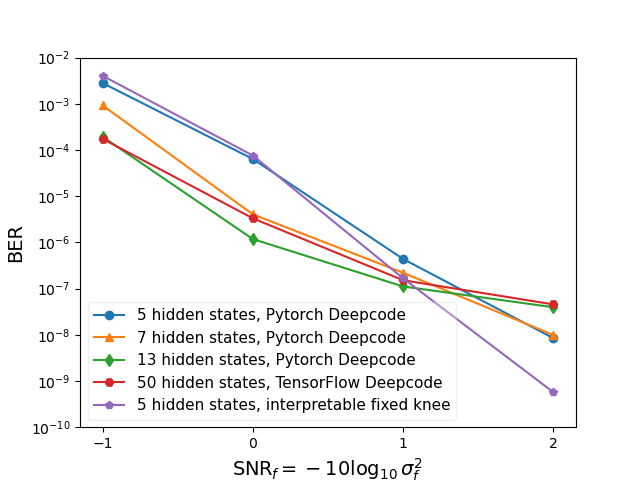}
    \caption{BER performance vs. forward SNR, different dimensions (number of hidden states in RNNs),  noiseless feedback}
    \vspace{-6mm}
    \label{fig:ber}
\end{figure}

The term ``dimension'' here refers to the number of hidden states in both the encoder and decoder. We tried two known model reduction techniques (Appendix \ref{apx: dimension reduction}), but their BER performance was much poorer than directly re-training models of smaller dimension. {In Deepcode's linear combination with $50$ hidden states (which generate parity bits), only around $13$ out of the $50$ trained weights had significant absolute values}; we thus pursued training models of dimension $\leq 13$ directly using the following technique: we initially trained Deepcode with reduced dimension using a block length of $100$ without power allocation and then re-trained with a block length of $50$ with power allocation inspired by \cite{shao2023attentioncode}. The corresponding performance is shown in Fig. \ref{fig:ber}. The parameters are trained at various forward SNRs, defined as $-10\log_{10}\sigma_f^2$. It turns out that a reduced dimension can approximate the performance of Deepcode with $50$ hidden states quite effectively. When the $\text{SNR}_f$ is low, larger dimension codes exhibit better performance. Conversely, when the $\text{SNR}_f$ is high,  smaller dimensions perform better. This may occur because training a large dimension becomes challenging in the presence of a small number of errors (small BER, small BCE). We expand on this in Section \ref{interpretation}.

\subsection{Pruning}
We observed that in the trained parameters, some values have significantly smaller  magnitude than others. Our next approach was simple: to prune (set to zero) the weights with small absolute values.
{It turns out that pruning smaller weights (up to a certain amount) has minimal impact on the model's BER performance, while yielding a more compact and efficient model} (see Appendix \ref{apx: pruning}).


In the following sections, \textit{``reduced Deepcode ($N_h$)''}  refers to the model obtained by retraining from scratch with $N_h$ hidden states in PyTorch. This process includes pruning the encoder and retraining the decoder. To ensure validity, we generate scatter plots illustrating the relationship between parity bits and phase 1 noises (Appendix \ref{apx: parity and phase1}), demonstrating consistency with Deepcode \cite{kim2018deepcode}. We focus on the simplest reduced Deepcode ($5$) later.


\section{Opening the black box}
In this section, we open the black box  of Deepcode's encoder by looking first at the influence length of the learned encoder RNNs based on different input perturbations, and then dig further to understand the actual function learned through the use of scatter plots and regression. 

\subsection{Influence Length}
For a specific input $\beta$, we define the expected $L_1$ difference of parity bits at time instant $i$ as $L_{i, \beta, \Delta} = \mathbb{E} \lVert f(\mathbf{P}_i) - f(\mathbf{P}_i^{(\beta, \Delta)}) \rVert_1$. Here, $\mathbf{P}_i$ is the RNN input, and the vector $\mathbf{P}_i^{(\beta, \Delta)}$ is obtained by perturbing the element $\beta$ in the vector $\mathbf{P}_i$ by $\Delta$. E.g., flipping the message bit results in  $\mathbf{P_i}^{(b_i, \Delta)} = \left[b_i\bigoplus1, n_i, n_{i-1,1}, n_{i-1,2}\right]$, where $\Delta = 1 \in \mathbb{F}_2$ or perturbing one of the noise components gives $\mathbf{P}_i^{(n_i,\Delta)} = \left[b_i, n_i - \Delta, n_{i-1,1}, n_{i-1,2}\right]$, for $\Delta \in \mathbb{R}$ the perturbation. 

The \textit{influence length} of input $\beta$ captures the number of time steps over which a change in the input affects the parity bits. Given a small value $\delta$ (set at $0.05$) and a perturbation of the input at position $t$ (randomly chosen as $t = 5$), the influence length is defined as: 
\begin{equation}
    \mathcal{L}_{\beta,\Delta} = \sum_{i = t}^K \mathbb{I}\left( L_{i, \beta, \Delta} > \delta \max_{k\in\{t,\ldots,K\}}{L_{k,\beta, \Delta}} \right).
\end{equation}

For each model, $\mathcal{L}_{b_i, \Delta}$ (Table. \ref{table:inf_b} of Appendix \ref{apx: influence}) remains constant, while the influence length for the noise (Table. \ref{table:inf_n} of Appendix \ref{apx: influence}) increases with $\Delta$, but eventually levels off. Together with BER plots, we conclude that longer influence lengths are effective in addressing rare events when the noises become extremely large. Based on our experimental maximum influence lengths, it seems unnecessary to use $50$ hidden states (as in the original Deepcode) to achieve comparable influence lengths and similar BER performance. As the noises follow a Gaussian distribution, the probability of exceeding $3\sigma_f$ is low. Consequently, we establish our interpretation based on small $\Delta$ deviations $\mathcal{L}_{b_i,\Delta} = \mathcal{L}_{n_i,\Delta} = 2$ and $\mathcal{L}_{n_{i,1},\Delta} = \mathcal{L}_{n_{i,2},\Delta} = 1$.

\subsection{Nonlinear Dynamics}
We now further peel open the black box and look at the specific learned RNN functions directly. 
 The RNN of Deepcode's encoder (original or reduced) is a discrete nonlinear dynamical system $\mathbf{h}_{i} = \tanh(\mathbf{W_{hp}}\mathbf{P}_{i} + \mathbf{W_{hh}}\mathbf{h}_{i-1}+ \mathbf{b})$ where $\mathbf{W_{hp}}, \mathbf{W_{hh}}, \mathbf{b}$ are the learned parameters. After pruning, the learned RNN parameters in the reduced Deepcode ($5$) have a special structure:
\begin{align}
   h_{i,p} = \begin{cases}
    q_{p,1}(b_i, n_{i}), & \text{if } p \in \{1, 2, 3\} \\
    q_{p,2}(n_{i-1,1}, n_{i-1,2},\mathbf{h}_{i-1}), & \text{if } p \in \{4, 5\}
\end{cases}
\end{align}
where $h_{i,p}$ represents the $p$-th element of hidden state $\mathbf{h}_{i}$.   In particular, each of the $5$ hidden states is either a non-recurrent $q_{p,1}$  function only of the message bits and phase 1 noise, or is a recurrent $q_{p,2}$ function only of the phase 2 noises and past hidden state.
The parity bits are linear combinations of the hidden state elements $c_{i,j} = \sum_{p=1}^{3}\alpha_{p,j} q_{p,1}(b_i, n_{i}) +  \sum_{p=4}^{5} \alpha_{p,j}q_{p,2}(n_{i-1,1}, n_{i-1,2}, \mathbf{h}_{i-1})$, where $\alpha_{p,j}$ are the learned coefficients. We represent the non-recurrent component of $c_{i,1}$ as $q_{1}$ and that of $c_{i,2}$ as $q'_{1}$.


\textit{Approximating the non-recurrent terms}: 
Inspired by Deepcode \cite{kim2018deepcode}, for the non-recurrent terms, we consider the scatter-plot (uniform message bits, phase 1 noises Gaussian) of functions $q_{1}$ and $q'_{1}$ (Fig. \ref{fig:pwl} of the Appendix \ref{apx: pwl}). They show that functions $q_1$ and $q_1'$ appear piecewise linear (PWL)  for each of the two possible message bits $b_i=0$ and $b_i=1$ (the original TensorFlow Deepcode has a very similar shape), and hence these functions may be well approximated by two segments. {We will use the indicator function $\mathbb{I}$ to represent the segments by adjusting the slopes later.}

\begin{figure}[ht]
    \vspace{-5mm}
    \centering
    \includegraphics[width=0.45\textwidth]{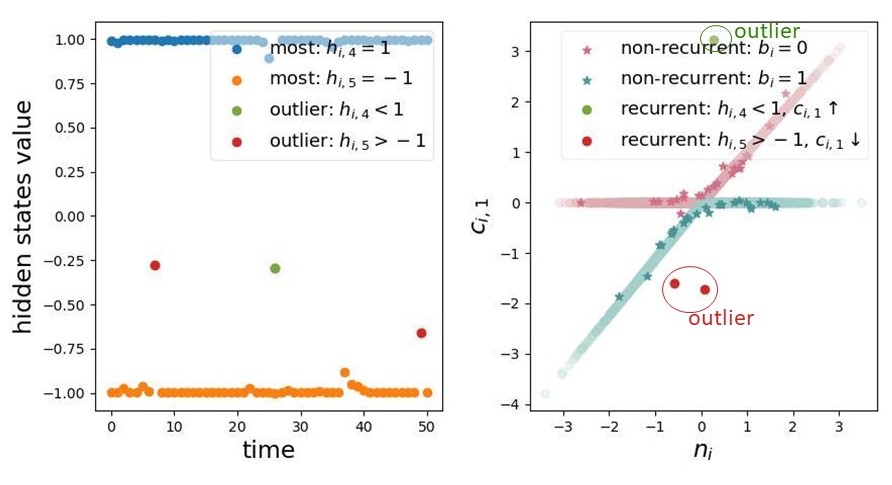}
    \caption{\small Outlier values of hidden states (left) and their impact on parity bit $c_{i,1}$ (right). {The outliers cause deviations in the parity bits from the regular values in the right figure.}}
    \vspace{-5mm}
    \label{fig:outlier}
\end{figure}

\textit{Understanding the recurrent terms through outlier analysis}: Unlike the non-recurrent terms, both $h_{i,4}$ and $h_{i,5}$ depend on previous inputs.  Consequently, we visualize the trajectory of the hidden states over time. {We plot the black box function over Gaussian noises and true bits.} From Fig. \ref{fig:outlier}, we observe that $h_{i,4}$ predominantly takes the value of $1$, while $h_{i,5}$ is mostly $-1$.  Notably, $h_{i,4}$ has outliers when the past message bit is $0$ (similarly $h_{i,5}$ has outliers when the bit is $1$), and the noises are unusually large. In most cases, the values of $h_{i,4}$ and $h_{i,5}$ cancel each other out when forming the parity bits. However, outliers significantly impact both parity bits, signaling and enabling error correction, as further interpreted in Section \ref{interpretation}. 




\section{Interpretable model of a Deepcode-like encoder and decoder}\label{interpretation}
Having built up an understanding of the learned encoder in the last sections, we now present an interpretable model built on the reduced models. 
{By \textit{interpretable} model we mean succinct non-linear expressions for the encoder $f$ and decoder $g$ that yield BER that closely resembles that of Deepcode.} We consider the reduced Deepcode ($5$) that captures the essential dynamics of Deepcode. The ideas carry over but become more complex as the number of hidden states increases. The interpretable models confirm that the encoder tries to use the feedback to assist the decoder in error correction. 
\subsection{Encoder Interpretation} \label{encoder interpretation} 

\subsubsection{Reduced Deepcode ($5$)}
We suggest the following basic interpretable model with $5$ hidden states:
\begin{align}
c_i & = 2b_{i} -1 \text{ (Deepcode's phase 1 -- uncoded transmission)} \\
c_{i, 1} &= \eqnmarkbox[red]{parity1}{e_1n_{i}\mathbb{I}(-(2b_i-1)n_{i})}  \eqnmarkbox[cyan]{parity1r}{- e_2h_{i,4} - e_2 h_{i,5}} \label{eq:intparity1}\\
c_{i, 2} &= \eqnmarkbox[red]{parity2}{- e_1n_{i}\mathbb{I}(-(2b_i-1)n_{i})} \eqnmarkbox[cyan]{parity2r}{- e_2h_{i,4} - e_2 h_{i,5}} \label{eq:intparity2}
\end{align}
where $e_1$ and $e_2$ are learned coefficients. 
\annotate[yshift=0.5em, xshift = 0.6em]{left}{parity1}{non-recurrent}
\annotate[yshift=0.5em, xshift = 0.6em]{right}{parity1r}{outlier analysis}


\textit{Analysis}: In the ``\textcolor{red}{red}'' portion, the parity bits $c_{i,1}$ and $c_{i,2}$ are energy efficient:  if the transmitted message bit is $b_i = 0$, and the noise in the first phase added to $b_i$ is negative, then the parity bits will not send new information about $n_{i}$ (not needed as a binary detection would yield the correct estimated bit). Otherwise, the parity bits contain a scaled version of $n_{i}$. The case is similar for $b_i=1$ and positive phase 1 noises.
A similar interpertation was obtained in \cite{kim2018deepcode}.  What is new with respect to the limited interpretation of \cite{kim2018deepcode} are the ``\textcolor{cyan}{blue}'' portions in 
 \eqref{eq:intparity1} and \eqref{eq:intparity2}. From our earlier studies, we know that feedback noises affect the outputs for 1 time unit in reduced Deepcode ($5$), i.e. $\mathcal{L}_{n_{i,1},\Delta} =\mathcal{L}_{n_{i,2},\Delta} = 1$.  {Therefore, $h_{i,4}$ and $h_{i,5}$ convey information from the last time step, as follows}: 
\begin{align}
 \text{If } b_{i-1}&=0, \text{ then } h_{i,5}=-1 \text{ and }\\
 h_{i,4} &= \tanh{(-k_1n_{i-1} + k_2n_{i-1, 1} - k_3n_{i-1, 2} + k_4)}  \label{eq:h4}\\
 \text{If } b_{i-1}&=1, \text{ then }  h_{i,4} =  1 \text{ and }\\
 h_{i,5} &=  \tanh{(-k_1n_{i-1} + k_2n_{i-1, 1} \label{eq:h5}
 - k_3n_{i-1, 2}- k_4)} 
 \end{align}
where $k_1$, $k_2$, $k_3$, and $k_4$ are learned coefficients. 

\textit{Analysis of the blue portion}: The blue portion serves as the error correction for the previous bit, and its value depends on the phase 1 and phase 2 noises from the last time step.

\begin{figure}[ht]
    \centering
    \includegraphics[width=0.45\textwidth]{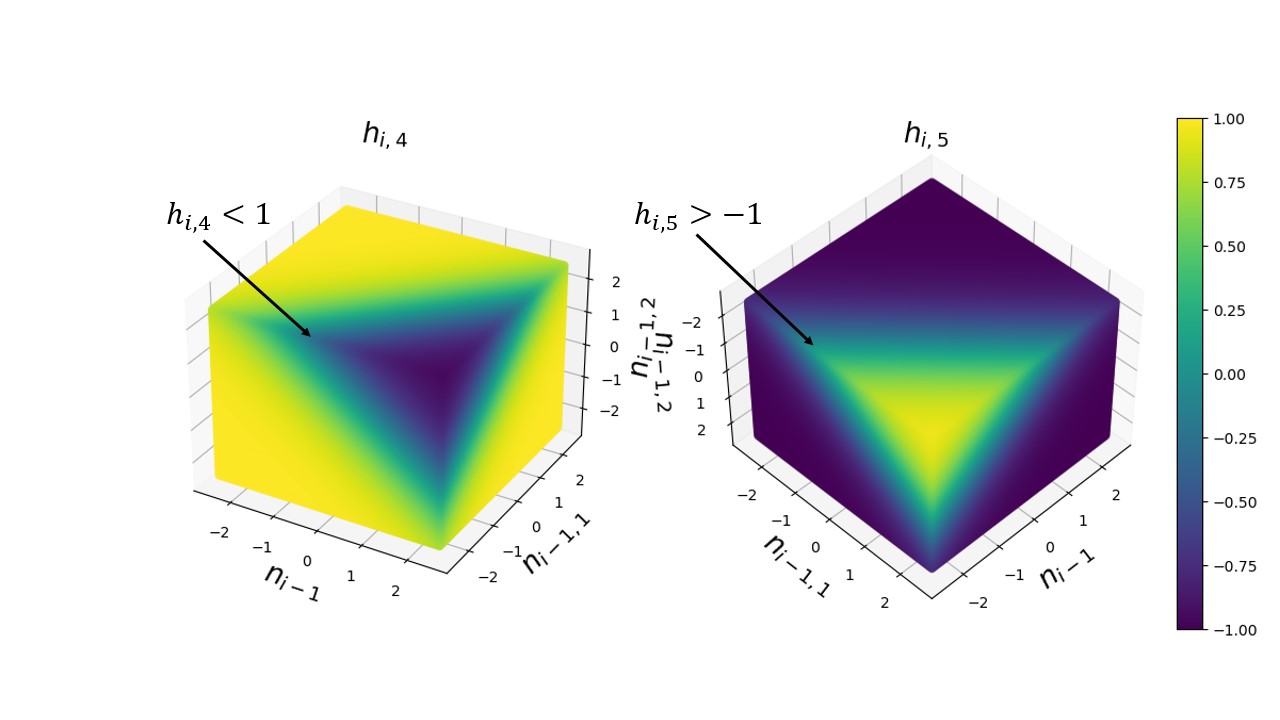}
    \caption{Outlier analysis of the ``\textcolor{cyan}{blue}'' portions in 
 \eqref{eq:intparity1} and \eqref{eq:intparity2}.}
    \vspace{-3mm}
    \label{fig:h value}
\end{figure}

$\bullet$ \textit{Without Outlier}: In small noise scenarios (as depicted in Fig. \ref{fig:h value}, yellow for $h_{i,4}$ or purple for $h_{i,5}$), $h_{i,4}$ ($1$) and $h_{i,5}$ ($-1$) cancel each other out in the parity functions.

    $\bullet$ \textit{With Outlier}: For purple values in the left and yellow on the right of Fig. \ref{fig:h value} we have outliers:

    \begin{itemize}
        \item[$-$] $h_{i,4}$:  outliers (value $<1$) occur when the message bit $b_{i-1}=0$, phase 1 noise $n_{i-1}$ is too positive, and phase 2 noise $n_{i-1,1}$ is too negative, or $n_{i-1,2}$ is too positive.
        \item[$-$] $h_{i,5}$: the situation is analogous to above, with $b_{i-1}=1$, and with the noise signs and magnitudes reversed.
    \end{itemize}
    %



\subsubsection{Reduced Deepcode ($7$)} {The ``enhanced'' interpretable encoder (for $N_h = 7$ hidden states) is designed to address more rare events and improve overall performance relative to the $N_h=5$ interpretable encoder. It introduces two more hidden states $h_{i, 6}$ and $h_{i,7}$ to match the longer observed influence lengths. The details are in Appendix \ref{7 hidden states}.}

\subsection{Decoder Interpretation}

 In the decoding process, we construct a simple non-linear code for $5$ hidden states using the symmetry of the parity bits.  Our decoder estimates the message bits as:
\begin{align}
O_{i,j} & = \tanh({d_{j,1}}y_i - {d_{j,2}}{(y_{i, 1}- y_{i, 2})} \\
& - {d_{j,3}}{(y_{i+1, 1} + y_{i+1, 2})} 
 + {d_{j,4}}) \\
& = \tanh(d_{j,1} \bar{c}_i + \eqnmarkbox[orange]{node4}{d_{j,1}n_i - d_{j,2}(\bar{c}_{i, 1}- \bar{c}_{i, 2})} \\ \label{eq: decoding1}
& \eqnmarkbox[teal]{node5}{- d_{j,2}(n_{i,1}-n_{i,2}) - d_{j,3}(\bar{c}_{i+1, 1} + \bar{c}_{i+1, 2})} \\[3mm]
& - d_{j,3}(n_{i+1,1}+n_{i+1,2}) + d_{j,4})  \\
\hat{b}_{i} & = \begin{cases}0 & \text{if } D_i < 0.5, \\ 1 & \text{if } D_i \geq 0.5. \end{cases} \quad D_i  = \sigma\left(\sum_{j=1}^{N_l=5}l_jO_{i,j}\right) \label{eq: decoding3}
\end{align}
where $\sigma$ represents the sigmoid function, and all  $\mathbf{d}$ and $\mathbf{l}$ values are learned parameters, and post power-allocation normalized parity bits are $\bar{c}_i$, $\bar{c}_{i,1}$ and $\bar{c}_{i,2}$. A linear combination of $N_l = 5$ decoding results is used in the final decision.
\annotate[yshift=0.5em, xshift = 1.2em]{right}{node4}{eliminate $n_i$}
\annotate[yshift= 0.1em, xshift = 2em]{below, label above}{node5}{error correction}

{Here, we assume the positivity of learned parameters in the encoder and decoder, except for the $d_{j,4}$ bias; other cases are addressed in symmetrically equivalent codes.}



\textit{Analysis}: the ``\textcolor{orange}{orange}'' (``\textcolor{teal}{teal}'') portion is used to eliminate phase 1 (2) noise in  \eqref{eq:intparity1} and \eqref{eq:intparity2}.
\begin{itemize}
    \item \textit{Phase 1 Noises}: The ``\textcolor{orange}{orange}'' portion in \eqref{eq: decoding1} eliminates the noises added in the first phase by subtracting the current parity bits.
    \item \textit{Phase 2 Noises}: The ``\textcolor{teal}{teal}'' portion in \eqref{eq: decoding1} eliminates phase 2 noises by adding future parity bits. Without outliers, the sum is about 0. With outliers, error correction is demonstrated in Table~\ref{decoder:h}:
    \begin{itemize}
        \item $h_{i+1,4}$($<1$): Consider the scenario where the message bit $b_i=0$; we want $O_{i,j}$ to be negative for correct detection. The outliers in $h_{i+1,4}$ will push the decoding value of $O_{i,j}$ to a negative value:
        \begin{equation}
            \eqnmarkbox[lightgray]{error}{- d_{j,2}(n_{i,1}^{\textcolor{red}{\downarrow}}-n_{i,2}^{\textcolor{red}{\uparrow}})}^{\textcolor{red}{\big\uparrow}} \eqnmarkbox[gray]{error}{- d_{j,3}(\bar{c}_{i+1, 1}^{\textcolor{red}{\uparrow}} + \bar{c}_{i+1, 2}^{\textcolor{red}{\uparrow}})}^{\textcolor{red}{\big\downarrow}}
        \end{equation}

        \item $h_{i+1,5}$($>-1$): For $b_i=1$, the situation is reversed, and $O_{i,j}$ needs to be positive for correct decoding.
        \begin{equation}
            \eqnmarkbox[lightgray]{error}{- d_{j,2}(n_{i,1}^{\textcolor{red}{\uparrow}}-n_{i,2}^{\textcolor{red}{\downarrow}})}^{\textcolor{red}{\big\downarrow}} \eqnmarkbox[gray]{error}{- d_{j,3}(\bar{c}_{i+1, 1}^{\textcolor{red}{\downarrow}} + \bar{c}_{i+1, 2}^{\textcolor{red}{\downarrow}})}^{\textcolor{red}{\big\uparrow}}
        \end{equation}
    \end{itemize}
    
\end{itemize}

\begin{table}[htbp]
\centering
\begin{tabular}{|l|l|l|l|l|l|l|l|}
\hline
$b_i$ &$n_{i}$ &$n_{i,1}$ & $n_{i,2}$  & $h_{i+1,4}$ & $h_{i+1,5}$ & $\bar{c}_{i+1,j}$  & $O_{i,j}$ \\ \hline
$0$ &$+$          & $-$          & $+$         & $< 1$ & $-1$        &$\uparrow$   & $\downarrow$     \\ \hline
$1$ &$-$          & $+$         &$-$         & $1$            & $>-1$ & $\downarrow$  & $\uparrow$ \\ \hline
\end{tabular}
\label{h1}
\caption{\label{decoder:h}Error Control}
\vspace{-2mm}
\end{table}




\subsection{Training}

Our interpretable encoder and decoder, based on $5$ hidden states in equations \eqref{eq:intparity1}, \eqref{eq:intparity2}, \eqref{eq: decoding1}, and  \eqref{eq: decoding3}, along with power allocation, have parameters ($\mathbf{e}$, $\mathbf{k}$, $\mathbf{d}$, $\mathbf{w}$, $\mathbf{a}$) that we learn rather than use to approximate Deepcode's outputs. During training, we focus on a block length of $K=50$. There are a total of $43$ parameters in our interpretable model, with $12$ of them ($\mathbf{w}$ and $\mathbf{a}$) associated with power allocation. We optimize all parameters to minimize BCE at different forward SNRs.

 Table. \ref{inter} presents the BER performance when matching different interpretable / original encoders / decoders of Deepcode. The results indicate that the interpretable encoder and decoder perform comparably to the reduced code. The BER performance of the interpretable model, based on 5 hidden states across different $\text{SNR}_f$, is shown in Fig. \ref{fig:ber}. Our interpretable model exhibits better performance than the original Deepcode at high $\text{SNR}_f$ but experiences decreased performance at low $\text{SNR}_f$ due to what we believe is its limited influence length for capturing unusual sequences of noise events (e.g. longer sequences of phase 2 noises being unusually large). {Moreover, we experimentally demonstrate that our encoder with $7$ hidden states improves the BER performance by an order of magnitude, making it comparable to Tensorflow Deepcode} (Table. \ref{7 hidden states BER} in Appendix \ref{7 hidden states}). We are working on developing an $N_h=7$ interpretable decoder for our $N_h=7$ encoder. 
\begin{table}[t]
\centering
\begin{tabular}{|l|l|l|l|}
\hline
Encoder & Decoder & BER  $\text{SNR}_f$ $0$  & BER $\text{SNR}_f$ $2$  \\ \hline
original    & original            & $6.375e-05$    & $8.417e-09$            \\ \hline
interpret.  & original      & $7.616e-05$    &   $1.300e-09  $  \\ \hline
original  & interpret.       & $7.498e-05$    & $5.350e-09$    \\ \hline
interpret.  & interpret.         &   $7.595e-05$  &   $5.694e-10$    \\ \hline
\end{tabular}
\caption{\label{inter}BER Performance of interpretable models of dimension 5 (noiseless feedback).}
\vspace{-5mm}
\end{table}

\subsection{Equivalent Codes}

The interpretable encoder and decoder, based on $5$ hidden states as discussed earlier, provides one particular example of possible codes. There are eight distinct equivalent combinations involving (1) positive or negative signs of components in the parity bits, and (2) the value selections in $h_{i,4}$ and $h_{i,5}$, which can be either $1$ or $-1$ (details are in Appendix \ref{apx: equivalent}). We verified (Appendix \ref{apx: equivalent} Table \ref{symmetry}) that all $8$ interpretable models yield comparable performance in terms of BER.


\subsection{Noisy Feedback}
After analyzing noiseless feedback, we expand our study to include noisy feedback. In the case of noisy feedback, reduced Deepcode ($5$) has a structure similar to that in the noiseless feedback, and all interpretation steps follow:


\begin{itemize}
    \item PWL approximation: from the scatter points we observe that the knee point shifts as the feedback SNR ($\text{SNR}_{fb}$) decreases (Appendix \ref{apx: pwl} Fig. \ref{fig:pwl_noisy}). Based on these observations, we extend our interpretable model to incorporate varying knee points and modify the non-recurrent part of the encoder $(n_{i}+\tilde{n}_{i})\mathbb{I}(-(2b_i-1)(n_{i}+\tilde{n}_{i}))$ slightly:
\begin{align}
 &\text{If } b_{i}=0, (n_{i}+\tilde{n}_{i} +\lambda_1)\mathbb{I}(-(2b_i-1)(n_{i}+\tilde{n}_{i} +\lambda_1))\\
 & \text{If } b_{i}=1, (n_{i}+\tilde{n}_{i} - \lambda_2)\mathbb{I}(-(2b_i-1)(n_{i}+\tilde{n}_{i} - \lambda_2))
 \end{align}
where $\lambda_1$ and $\lambda_2$ are learned parameters. 
    \item Outlier analysis: The number of outliers increases as the feedback SNR ($\text{SNR}_{fb}$) decreases. (Appendix \ref{apx: noisy feedback} Fig. \ref{fig:outlier_noisy}).
\end{itemize}

The BER performance is shown in Appendix \ref{apx: noisy feedback} Fig. \ref{fig:noisy ber}. With noisy feedback, the interpretable model with varying knee points performs as well as the $5$ hidden states Deepcode and slightly outperforms the interpretable model with fixed knee points. This suggests that the varying knee points adapt to the feedback noises and attempt to mitigate their impact.

\section{Conclusions}


We presented an interpretable model for the RNN-based Deepcode.  We demonstrated the impact of feedback on decoding through outliers and showed that our interpretable model performs comparably to the original Deepcode with significantly fewer parameters, both in noiseless and noisy feedback scenarios. Notably, it outperforms Deepcode at high forward SNR. However, at low $\text{SNR}_f$, our interpretable model exhibits reduced performance, suggesting that the short influence length of our interpretable model limits error correction. Future research may focus on exploring the interpretation of models with longer influence lengths, and on identifying an algorithm for the construction of (analytical) feedback codes with a given influence length.

\clearpage
\bibliographystyle{IEEEtran}
\bibliography{reference}

\begin{thebibliography}{10}
\providecommand{\url}[1]{#1}
\csname url@samestyle\endcsname
\providecommand{\newblock}{\relax}
\providecommand{\bibinfo}[2]{#2}
\providecommand{\BIBentrySTDinterwordspacing}{\spaceskip=0pt\relax}
\providecommand{\BIBentryALTinterwordstretchfactor}{4}
\providecommand{\BIBentryALTinterwordspacing}{\spaceskip=\fontdimen2\font plus
\BIBentryALTinterwordstretchfactor\fontdimen3\font minus \fontdimen4\font\relax}
\providecommand{\BIBforeignlanguage}[2]{{%
\expandafter\ifx\csname l@#1\endcsname\relax
\typeout{** WARNING: IEEEtran.bst: No hyphenation pattern has been}%
\typeout{** loaded for the language `#1'. Using the pattern for}%
\typeout{** the default language instead.}%
\else
\language=\csname l@#1\endcsname
\fi
#2}}
\providecommand{\BIBdecl}{\relax}
\BIBdecl

\bibitem{kim2018deepcode}
H.~Kim, Y.~Jiang, S.~Kannan, S.~Oh, and P.~Viswanath, ``Deepcode: Feedback codes via deep learning,'' \emph{Advances in neural information processing systems}, vol.~31, 2018.

\bibitem{safavi2021deep}
A.~R. Safavi, A.~G. Perotti, B.~M. Popovic, M.~B. Mashhadi, and D.~Gunduz, ``Deep extended feedback codes,'' \emph{arXiv preprint arXiv:2105.01365}, 2021.

\bibitem{mashhadi2021drf}
M.~B. Mashhadi, D.~Gunduz, A.~Perotti, and B.~Popovic, ``Drf codes: Deep snr-robust feedback codes,'' \emph{arXiv preprint arXiv:2112.11789}, 2021.

\bibitem{shao2023attentioncode}
Y.~Shao, E.~Ozfatura, A.~Perotti, B.~Popovic, and D.~G{\"u}nd{\"u}z, ``Attentioncode: Ultra-reliable feedback codes for short-packet communications,'' \emph{IEEE Transactions on Communications}, 2023.

\bibitem{ozfatura2022all}
E.~Ozfatura, Y.~Shao, A.~G. Perotti, B.~M. Popovi{\'c}, and D.~G{\"u}nd{\"u}z, ``All you need is feedback: Communication with block attention feedback codes,'' \emph{IEEE Journal on Selected Areas in Information Theory}, vol.~3, no.~3, pp. 587--602, 2022.

\bibitem{kim2020deepcode}
H.~Kim, Y.~Jiang, S.~Kannan, S.~Oh, and P.~Viswanath, ``Deepcode and modulo-sk are designed for different settings,'' \emph{arXiv preprint arXiv:2008.07997}, 2020.

\bibitem{ben2020simple}
A.~Ben-Yishai and O.~Shayevitz, ``Simple modulo can significantly outperform deep learning-based deepcode,'' \emph{arXiv preprint arXiv:2008.01686}, 2020.

\bibitem{kim2023robust}
J.~Kim, T.~Kim, D.~Love, and C.~Brinton, ``Robust non-linear feedback coding via power-constrained deep learning,'' \emph{arXiv preprint arXiv:2304.13178}, 2023.

\bibitem{schalkwijk1966coding}
J.~Schalkwijk and T.~Kailath, ``A coding scheme for additive noise channels with feedback--i: No bandwidth constraint,'' \emph{IEEE Transactions on Information Theory}, vol.~12, no.~2, pp. 172--182, 1966.

\bibitem{shayevitz2011optimal}
O.~Shayevitz and M.~Feder, ``Optimal feedback communication via posterior matching,'' \emph{IEEE Transactions on Information Theory}, vol.~57, no.~3, pp. 1186--1222, 2011.

\bibitem{ben2017interactive}
A.~Ben-Yishai and O.~Shayevitz, ``Interactive schemes for the awgn channel with noisy feedback,'' \emph{IEEE Transactions on Information Theory}, vol.~63, no.~4, pp. 2409--2427, 2017.

\bibitem{chance2011concatenated}
Z.~Chance and D.~J. Love, ``Concatenated coding for the awgn channel with noisy feedback,'' \emph{IEEE Transactions on Information Theory}, vol.~57, no.~10, pp. 6633--6649, 2011.

\bibitem{mishra2023linear}
R.~Mishra, D.~Vasal, and H.~Kim, ``Linear coding for awgn channels with noisy output feedback via dynamic programming,'' \emph{IEEE Transactions on Information Theory}, 2023.

\bibitem{lian2018can}
M.~Lian, C.~H{\"a}ger, and H.~D. Pfister, ``What can machine learning teach us about communications?'' in \emph{2018 IEEE Information Theory Workshop (ITW)}.\hskip 1em plus 0.5em minus 0.4em\relax IEEE, 2018, pp. 1--5.

\bibitem{sussillo2013opening}
D.~Sussillo and O.~Barak, ``Opening the black box: low-dimensional dynamics in high-dimensional recurrent neural networks,'' \emph{Neural computation}, vol.~25, no.~3, pp. 626--649, 2013.

\bibitem{karpathy2015visualizing}
A.~Karpathy, J.~Johnson, and L.~Fei-Fei, ``Visualizing and understanding recurrent networks,'' \emph{arXiv preprint arXiv:1506.02078}, 2015.

\bibitem{choe2017probabilistic}
Y.~J. Choe, J.~Shin, and N.~Spencer, ``Probabilistic interpretations of recurrent neural networks,'' \emph{Probabilistic Graphical Models}, 2017.

\bibitem{Jain19}
S.~Jain and B.~C. Wallace, ``Attention is not explanation,'' in \emph{Proceedings of the 2019 Conference of the North American Chapter of the Association for Computational Linguistics: Human Language Technologies, {NAACL-HLT} 2019, Volume 1 (Long and Short Papers)}, 2019, pp. 3543--3556.

\bibitem{Wiegr}
S.~Wiegreffe and Y.~Pinter, ``Attention is not not explanation,'' in \emph{Proceedings of the 2019 Conference on Empirical Methods in Natural Language Processing and the 9th International Joint Conference on Natural Language Processing, {EMNLP-IJCNLP} 2019}, 2019, pp. 11--20.

\bibitem{devroye2022interpreting}
N.~Devroye, N.~Mohammadi, A.~Mulgund, H.~Naik, R.~Shekhar, G.~Tur{\'a}n, Y.~Wei, and M.~{\v{Z}}efran, ``Interpreting deep-learned error-correcting codes,'' in \emph{2022 IEEE International Symposium on Information Theory (ISIT)}.\hskip 1em plus 0.5em minus 0.4em\relax IEEE, 2022, pp. 2457--2462.

\bibitem{mulgund2022evaluating}
A.~Mulgund, R.~Shekhar, N.~Devroye, G.~Tur{\'a}n, and M.~{\v{Z}}efran, ``Evaluating interpretations of deep-learned error-correcting codes,'' in \emph{2022 58th Annual Allerton Conference on Communication, Control, and Computing (Allerton)}.\hskip 1em plus 0.5em minus 0.4em\relax IEEE, 2022, pp. 1--8.

\bibitem{devroye2023interpreting}
N.~Devroye, A.~Mulgund, R.~Shekhar, G.~Tur{\'a}n, M.~{\v{Z}}efran, and Y.~Zhou, ``Interpreting training aspects of deep-learned error-correcting codes,'' in \emph{2023 IEEE International Symposium on Information Theory (ISIT)}.\hskip 1em plus 0.5em minus 0.4em\relax IEEE, 2023, pp. 2374--2379.

\bibitem{mulgund2023decomposing}
A.~Mulgund, N.~Devroye, G.~Tur{\'a}n, and M.~{\v{Z}}efran, ``Decomposing the training of deep learned turbo codes via a feasible map decoder,'' in \emph{2023 12th International Symposium on Topics in Coding (ISTC)}.\hskip 1em plus 0.5em minus 0.4em\relax IEEE, 2023, pp. 1--5.

\bibitem{jiang2019turbo}
Y.~Jiang, H.~Kim, H.~Asnani, S.~Kannan, S.~Oh, and P.~Viswanath, ``Turbo autoencoder: Deep learning based channel codes for point-to-point communication channels,'' \emph{Advances in neural information processing systems}, vol.~32, 2019.

\bibitem{lall2002subspace}
S.~Lall, J.~E. Marsden, and S.~Glava{\v{s}}ki, ``A subspace approach to balanced truncation for model reduction of nonlinear control systems,'' \emph{International Journal of Robust and Nonlinear Control: IFAC-Affiliated Journal}, vol.~12, no.~6, pp. 519--535, 2002.

\bibitem{kleinberg2006algorithm}
J.~Kleinberg and E.~Tardos, \emph{Algorithm design}.\hskip 1em plus 0.5em minus 0.4em\relax Pearson Education India, 2006.

\end{thebibliography}

\clearpage
\newpage
\appendices

\section{Model Reduction} \label{apx: model reduction}
In this section, we provide more details on model reduction. 
\subsection{Dimension Reduction}\label{apx: dimension reduction}
We first attempted to reduce the dimension of Deepcode through two known methods. First, we tried the construction of an empirical balanced realization \cite{lall2002subspace}. The main idea is to project the nonlinear dynamics onto a lower-dimensional subspace while preserving most of the input-output relationships. However, our implementation of this method exhibited poor BER performance in low dimensions,  despite having an energy ratio approaching $1$. Second, we tried teacher-student training.  Here, we teach a reduced dimension version of  Deepcode with  fewer hidden states  using the knowledge from Deepcode with $N_h=50$ hidden states.  This method worked somewhat, but surprisingly, not as well as simply training a reduced dimension Deepcode directly. 
An interesting open question is how to find a more general approach to reduce the model dimension while preserving BER performance.

\subsection{Pruning parameter performance} \label{apx: pruning}

{In this subsection, we demonstrate the effects of pruning the RNN encoder's learned weights. We selectively set smaller parameters to zero based on their absolute values.}

\begin{figure}[ht]
    \centering
    \includegraphics[width=0.45\textwidth]{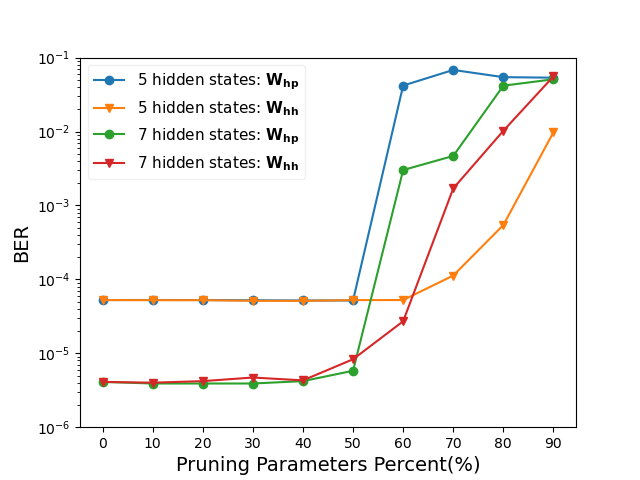}
    \caption{ BER Performance of Pruning in PyTorch Deepcode with $5$ and $7$ hidden states ($\text{SNR}_f = 0$, noiseless feedback)}
    \label{fig:pruning}
\end{figure}

{
Fig. \ref{fig:pruning} experimentally shows that for models with $5$ and $7$ hidden states, pruning 50\% of the parameters in the input-to-hidden weight matrix $\mathbf{W_{hp}}$ is the maximum tolerable amount; surpassing this threshold results in a dramatic increase in BER performance. For the hidden-to-hidden weight matrix $\mathbf{W_{hh}}$, the pruning limit is 60\% in models with $5$ hidden states and 40\% in those with $7$ hidden states. We hence choose to prune most aggressively, removing all weights that do not affect the BER performance.}


\section{Influence Length} \label{apx: influence}
In this section, we present the influence length when perturbing the noises or flipping the message bits. The influence length depends on the threshold $\delta$  selected, but the actual values for the same $\delta$ allow one to compare across models.

For the noises, as the perturbation value $\Delta \in \mathbb{R}$ increases, the influence length for each model levels out to a maximum value. From Table. \ref{table:inf_n}, we can see that the maxima of $\mathcal{L}_{n_i,\Delta}, \mathcal{L}_{n_{i,1},\Delta}$ and $\mathcal{L}_{n_{i,2},\Delta}$ are all equal to $6$
for TensorFlow Deepcode with $50$ hidden states. Interestingly, PyTorch Deepcode with only $7$ hidden states achieves a similar maximum influence length. We believe that a longer influence length is more effective in addressing rare events, especially when the noises are extremely large. This suggests that  50 hidden states may not be necessary -- 7 may suffice (in terms of BER performance). 
Furthermore, our interpretable model is constrained to $\mathcal{L}_{n_{i,1},\Delta}= \mathcal{L}_{n_{i,2},\Delta} = 1$, being able to correct only on ``abnormally large'' phase 2 noise in the previous time instance. This limitation affects our Bit Error Rate (BER) performance. A potential future direction is to extend the influence lengths for the interpretable model, which we surmise could lead to canceling error events in which a longer sequence of phase 2 corrective noises are all abnormally large. 

\begin{table}[H]
\centering
\begin{tabular}{|p{1.5cm}|p{0.6cm}|p{0.75cm}|p{0.75cm}|p{0.6cm}|p{0.75cm}|p{0.75cm}|}
\hline
\multirow{2}{*}{Model} & \multicolumn{3}{c|}{$\Delta = \pm 1$} & \multicolumn{3}{c|}{$\Delta = \pm 2$} \\
\cline{2-7}
& $\mathcal{L}_{n_i,\Delta}$ & $\mathcal{L}_{n_{i,1},\Delta}$ &  $\mathcal{L}_{n_{i,2},\Delta}$ & $\mathcal{L}_{n_i,\Delta}$ & $\mathcal{L}_{n_{i,1},\Delta}$ &  $\mathcal{L}_{n_{i,2},\Delta}$ \\ 
\hline
PTDeep ($5$)          & $2$          & $1$ &      $1$  & $2$          & $1$ &      $1$    \\ \hline
PTDeep ($7$)          & $3$         &$2$     &$3$ & $4$          & $3$ &      $3$   \\ \hline
PTDeep  ($50$)         & $3$         &$3$     &$3$  & $4$          & $3$ &      $3$      \\ \hline
TFDeep ($50$)         & $4$         &$3$     &$3$  & $5$          & $4$ &      $4$        \\ \hline
Interpretable          & $2$         &$1$     &$1$   & $2$          & $1$ &      $1$       \\ 
\hline
\multirow{2}{*}{Model} & \multicolumn{3}{c|}{$\Delta = \pm 3$} & \multicolumn{3}{c|}{$\Delta = \pm 5$} \\
\cline{2-7}
& $\mathcal{L}_{n_i,\Delta}$ & $\mathcal{L}_{n_{i,1},\Delta}$ &  $\mathcal{L}_{n_{i,2},\Delta}$ & $\mathcal{L}_{n_i,\Delta}$ & $\mathcal{L}_{n_{i,1},\Delta}$ &  $\mathcal{L}_{n_{i,2},\Delta}$ \\ 
\hline
PTDeep ($5$)          & $2$          & $2$ &      $2$  & $3$          & $2$ &      $2$    \\ \hline
PTDeep ($7$)          & $4$         &$3$     &$4$ & $4$          & $5$ &      $5$   \\ \hline
PTDeep  ($50$)         & $4$         &$3$     &$3$  & $4$          & $4$ &      $3$      \\ \hline
TFDeep ($50$)         & $5$         &$4$     &$4$  & $5$          & $4$ &      $5$        \\ \hline
Interpretable          & $2$         &$1$     &$1$   & $2$          & $1$ &      $1$       \\ 
\hline
\multirow{2}{*}{Model} & \multicolumn{3}{c|}{$\Delta = \pm 10$} & \multicolumn{3}{c|}{$\Delta = \pm 100/200$ } \\
\cline{2-7}
& $\mathcal{L}_{n_i,\Delta}$ & $\mathcal{L}_{n_{i,1},\Delta}$ &  $\mathcal{L}_{n_{i,2},\Delta}$ & $\mathcal{L}_{n_i,\Delta}$ & $\mathcal{L}_{n_{i,1},\Delta}$ &  $\mathcal{L}_{n_{i,2},\Delta}$ \\ 
\hline
PTDeep ($5$)          & $3$          & $2$ &      $2$  & $3$          & $2$ &      $2$    \\ \hline
PTDeep ($7$)          & $5$         &$6$     &$6$ & $5$          & $6$ &      $6$   \\ \hline
PTDeep  ($50$)         & $4$         &$5$     &$4$  & $5$          & $5$ &      $5$      \\ \hline
TFDeep ($50$)         & $6$         &$5$     &$6$  & $6$          & $6$ &      $6$        \\ \hline
Interpretable          & $2$         &$1$     &$1$   & $2$          & $1$ &      $1$       \\ \hline
\end{tabular}
\label{h2}
\caption{\label{table:inf_n}Influence Length of perturbing noises ($\text{SNR}_f = 0$, noiseless feedback). We denote PTDeep ($N_h$) for PyTorch Deepcode and TFDeep ($N_h$) for TensorFlow Deepcode with $N_h$ hidden states, respectively. Interpretable refers to our interpretable model with fixed knee points.}
\end{table}


For the message bits in Table. \ref{table:inf_b}, more hidden states appear to result in longer influence lengths, but not significantly longer.

\begin{table}[H]
\centering
\begin{tabular}{|p{2.3cm}|p{0.6cm}|}
\hline
Model  & $\mathcal{L}_{b_i,\Delta}$ \\ 
\hline
PTDeep ($5$)          & $2$               \\ \hline
PTDeep ($7$)          & $3$         \\ \hline
PTDeep  ($50$)         & $3$            \\ \hline
TFDeep ($50$)         & $4$              \\ \hline
Interpretable          & $2$               \\ 
\hline         
\end{tabular}
\caption{\label{table:inf_b}Influence Length of flipping message bits ($\text{SNR}_f = 0$, noiseless feedback).}
\end{table}

\section{Piecewise linear approximation of the non-recursive portion of the parity bits.} \label{apx: pwl}
In this section, we discuss the piecewise linear approximation in our analysis. 
The general methodology here is to perform a scatter plot and then fit each perceived learned function (for each input bit one would assume a different function) with a segmented least square, which is based on dynamic programming.  By carefully tuning the penalty for introducing new segments, we can attain an optimal piecewise linear approximation \cite{kleinberg2006algorithm}.  

\begin{figure}[ht]
    \centering
    \includegraphics[width=0.45\textwidth]{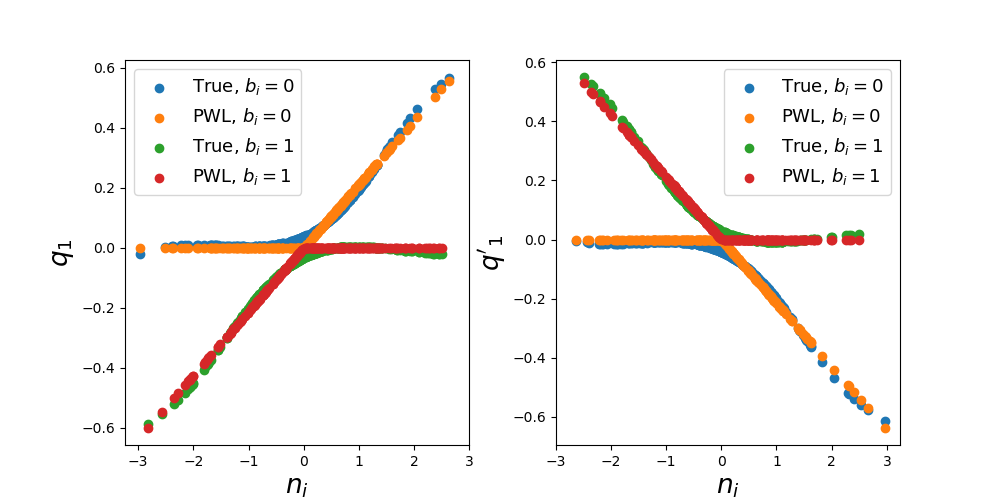}
    \caption{Piecewise linear approximation of non-recurrent functions (noiseless feedback)}
    \label{fig:pwl}
\end{figure}

In the noiseless feedback case in Fig. \ref{fig:pwl}, We set the knee point at the origin and perform piece-wise linear approximation to minimize the mean squared error (MSE). {For the flat portion, we ignore the slope since it is very small and set the value to $0$. On the steep side, we adjust the slope accordingly.}

\begin{figure}[ht]
    \centering
    \includegraphics[width=0.45\textwidth]{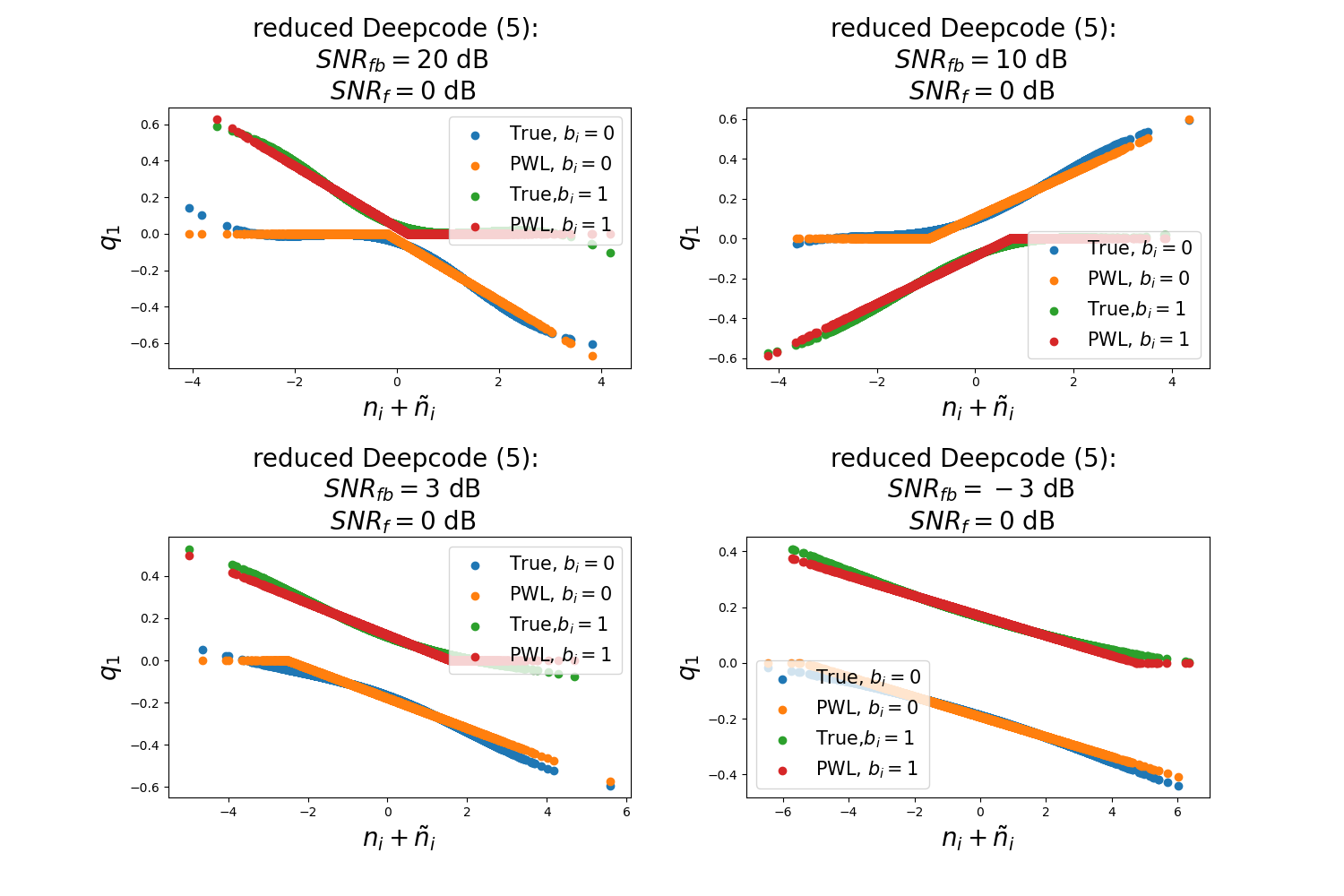}
    \caption{Piecewise linear approximation of non-recurrent functions (noisy feedback)}
    \label{fig:pwl_noisy}
\end{figure}

In the noisy feedback case, the scatter plots in Fig. \ref{fig:pwl_noisy} show that the knee points depend on the feedback noise variance. We again set the value to $0$ on the flat portion and adjust the slope and also the knee points accordingly.

\section{Encoder Interpretation: 7 hidden states} \label{7 hidden states}



In this section, we expand on the $7$ hidden states encoder. 
As we saw earlier, when we have $N_h=7$ hidden states the maximum influence of phase 2 noises on the parity bits has a length of $6$, and hence our interpretable encoder must capture these longer dependencies. This will also allow for more refined error correction when longer sequences of rare events occur, and noises in even the correction phase are unusually large.  

Through outlier analysis of the two new hidden states $h_{i,6}$ and $h_{i,7}$, we found that their values are primarily $1$ unless outliers are present. It is important to note that there are four types of sign combinations for phase 2 noises (e.g., both positive, both negative, one positive, and the other negative). At each time instant, only one of the $h_{i, j}$ values, where $j \in \{4, 5, 6, 7\}$, exhibits outliers, as they handle each case individually.

After a slight modification, our interpretable encoder for $7$ hidden states is as follows: 
\begin{align}
c_i  = & 2b_{i} -1 \\
c_{i, 1} = & e_1n_{i}\mathbb{I}(-(2b_i-1)n_{i}) - e_2h_{i,4} - e_2 h_{i,5}\\
& - e_3h_{i,6} + e_3h_{i,7}\\
c_{i, 2} = &- e_1n_{i}\mathbb{I}(-(2b_i-1)n_{i}) - e_2h_{i,4} - e_2 h_{i,5} \\
& - e_3h_{i,6} + e_3h_{i,7} 
\end{align}
where $h_{i,4}$ and $h_{i,5}$ correspond to \eqref{eq:h4} and  \eqref{eq:h5}, respectively. 

When outliers are absent, we aim for the values of $h_{i,6}$ and $h_{i,7}$ to offset each other. However, in the presence of outliers, they can function similarly to $h_{i,4}$ and $h_{i,5}$ to perform error correction. They are also used to extend the influence lengths. Based on these, the new $h_{i,6}$ and $h_{i,7}$ are introduced as:
\begin{align}
h_{i,6} & = \tanh\left(m_1n_{i-1, 1} + m_2n_{i-1, 2} + m_3h_{i-1,4} \right. \nonumber \\
& \quad \left. + m_4h_{i-1,7} + m_5\right) \\
h_{i,7} & = \tanh\left(-m_1n_{i-1, 1} - m_2n_{i-1, 2} - m_3h_{i-1,5} \right. \nonumber \\
& \quad \left. + m_4h_{i-1,6} + m_5\right)
\end{align}
where $\mathbf{m}$ values are all learned coefficients. 

Table. \ref{7 hidden states BER} shows BER performance when employing various interpretable and original encoders/decoders in Deepcode. Together with Table. \ref{inter}, the experiment shows that using a decoder with the same dimension of $5$, incorporating $7$ hidden states encoder,  significantly improves the BER performance. Moreover, our interpretable encoder effectively approximates the original encoder.


\begin{table}[t]
\centering
\begin{tabular}{|l|l|l|}
\hline
Encoder & Decoder & BER ($\text{SNR}_f$ $0$)    \\ \hline
original (7)    & original (7)            & $4.048e-06$             \\ \hline
original (7)    & original (5)            & $5.419e-06    $             \\ \hline
interpret. (7)  & original (7)      & $4.066e-06$     \\ \hline
interpret. (7) & original (5)        & $6.416e-06$  \\ \hline
\end{tabular}
\caption{\label{7 hidden states BER}BER Performance of interpretable encoder of dimension $7$ (noiseless feedback). The number of hidden states is in parentheses.}
\vspace{-5mm}
\end{table}

\section{Parity Bits Scatter Plot} \label{apx: parity and phase1}
In this section, we follow the idea first presented in \cite{kim2018deepcode} to generate the scatter plots for understanding the underlying encoder function.

\begin{figure}[htbp]
    \centering
    \includegraphics[width=0.45\textwidth]{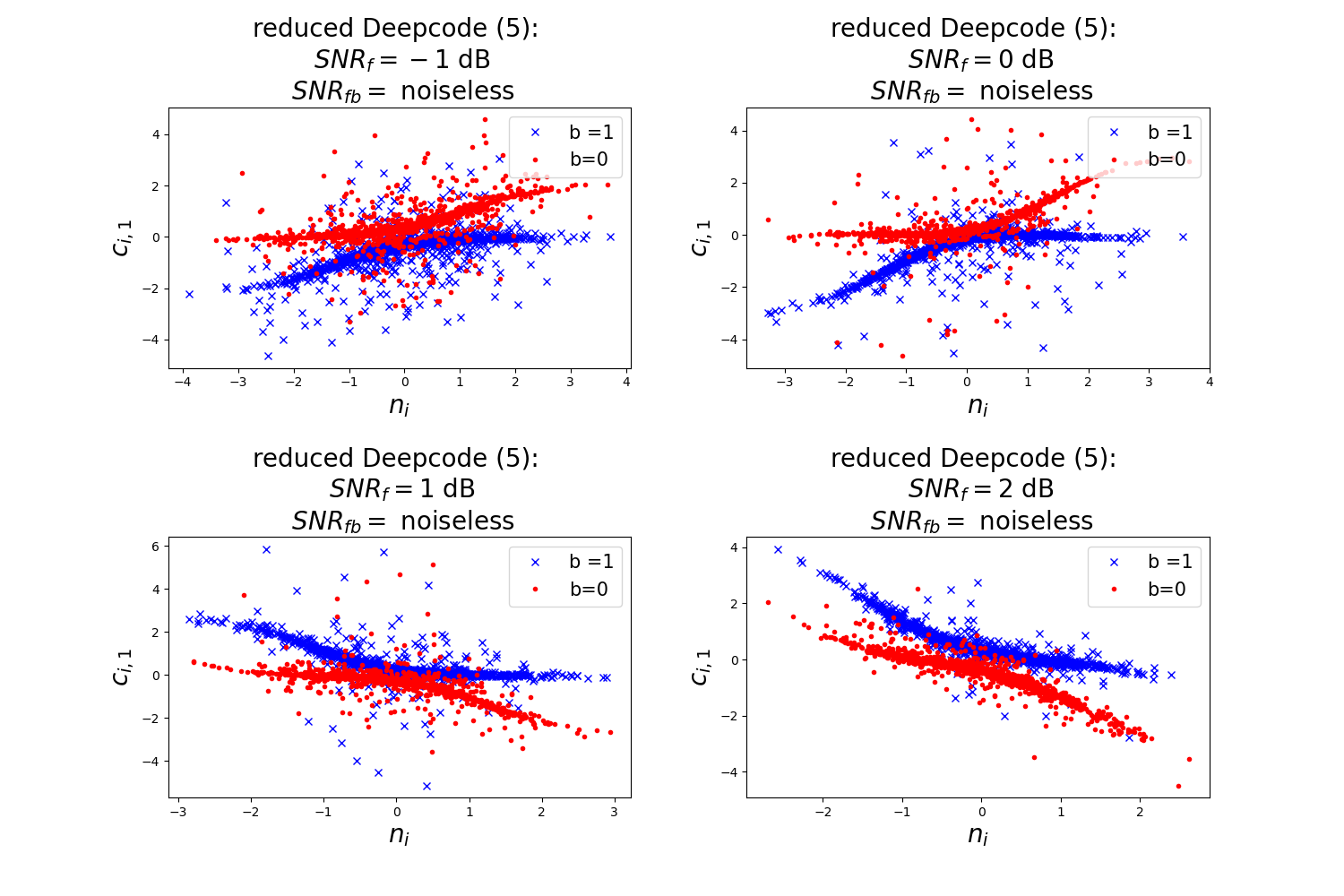}
    \caption{Reduced Deepcode ($5$): parity bit $c_{i,1}$ vs. forward noises in the first phase $n_i$}
    \label{fig:noiseless parity}
\end{figure}

\begin{figure}[htbp]
    \centering
    \includegraphics[width=0.45\textwidth]{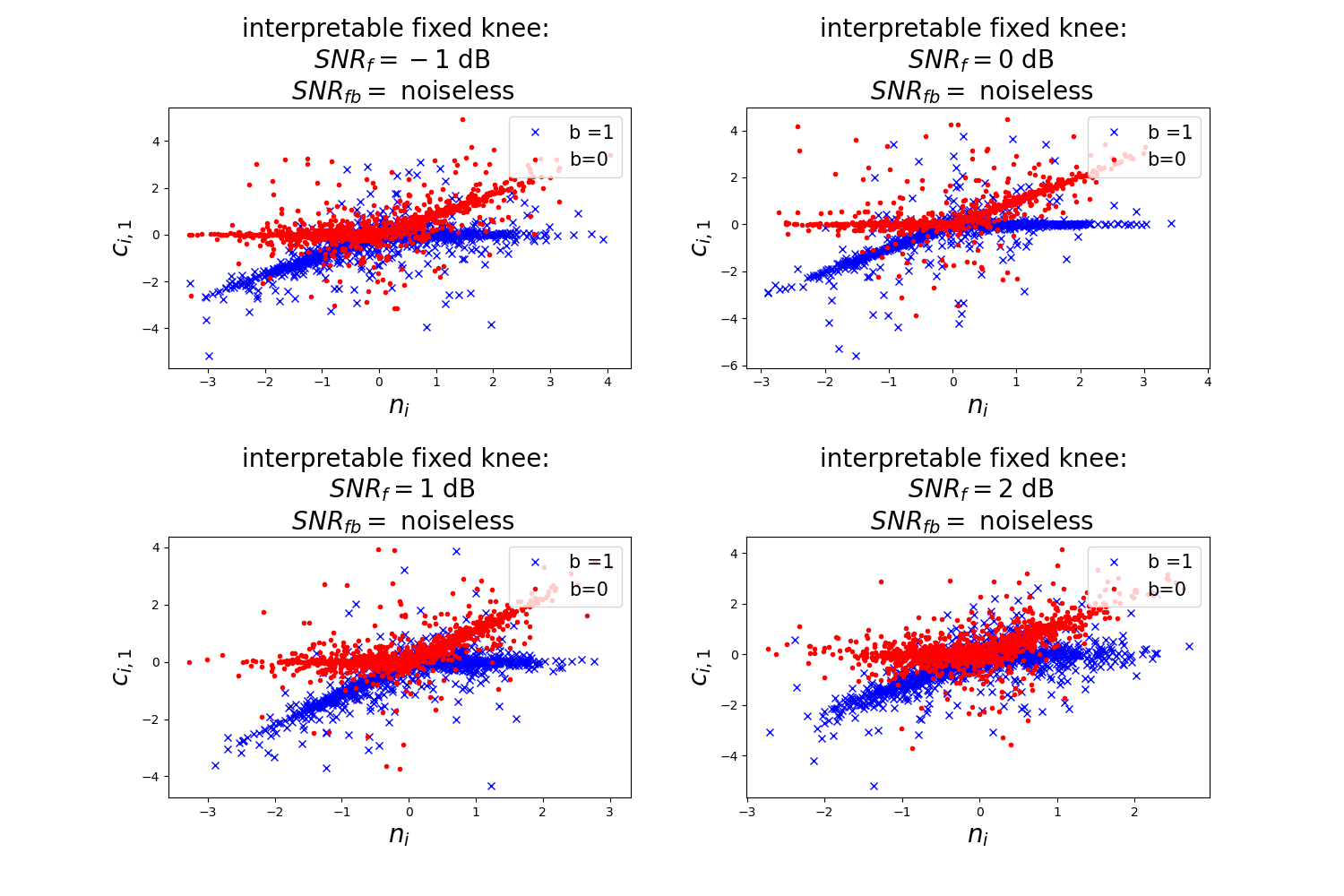}
    \caption{Interpretable fixed knee: parity bit $c_{i,1}$ vs. forward noises in the first phase $n_i$}
    \label{fig:noiseless parity inter}
\end{figure}

In the case of noiseless feedback, Figs. \ref{fig:noiseless parity} and \ref{fig:noiseless parity inter} show the evolving pattern of parity bits as the forward SNRs increase. Each scatter plot displays $2500$ sample points: $50$ samples for each $i$ where $i \in \{1, \ldots, 50\}$. The behavior of the parity bits always roughly has a flat portion and a sloped portion; whether they go up or down is more or less equivalent due to the symmetries. 

For reduced Deepcode ($5$), the parity bits shift from a non-linear ReLU-like pattern to a flatter linear function as the forward SNRs increase (Fig.~\ref{fig:noiseless parity}). Meanwhile, the interpretable model maintains the ReLU-like shape (Fig. \ref{fig:noiseless parity inter}). We hypothesize that our interpretable model outperforms Deepcode at high forward SNRs because there are fewer errors (at high SNRs) from which Deepcode can learn the correct shape. When the forward SNR is $-1$ dB, our interpretable model provides less error correction due to its limited influence length. 


\begin{figure}[htbp]
    \centering
    \includegraphics[width=0.45\textwidth]{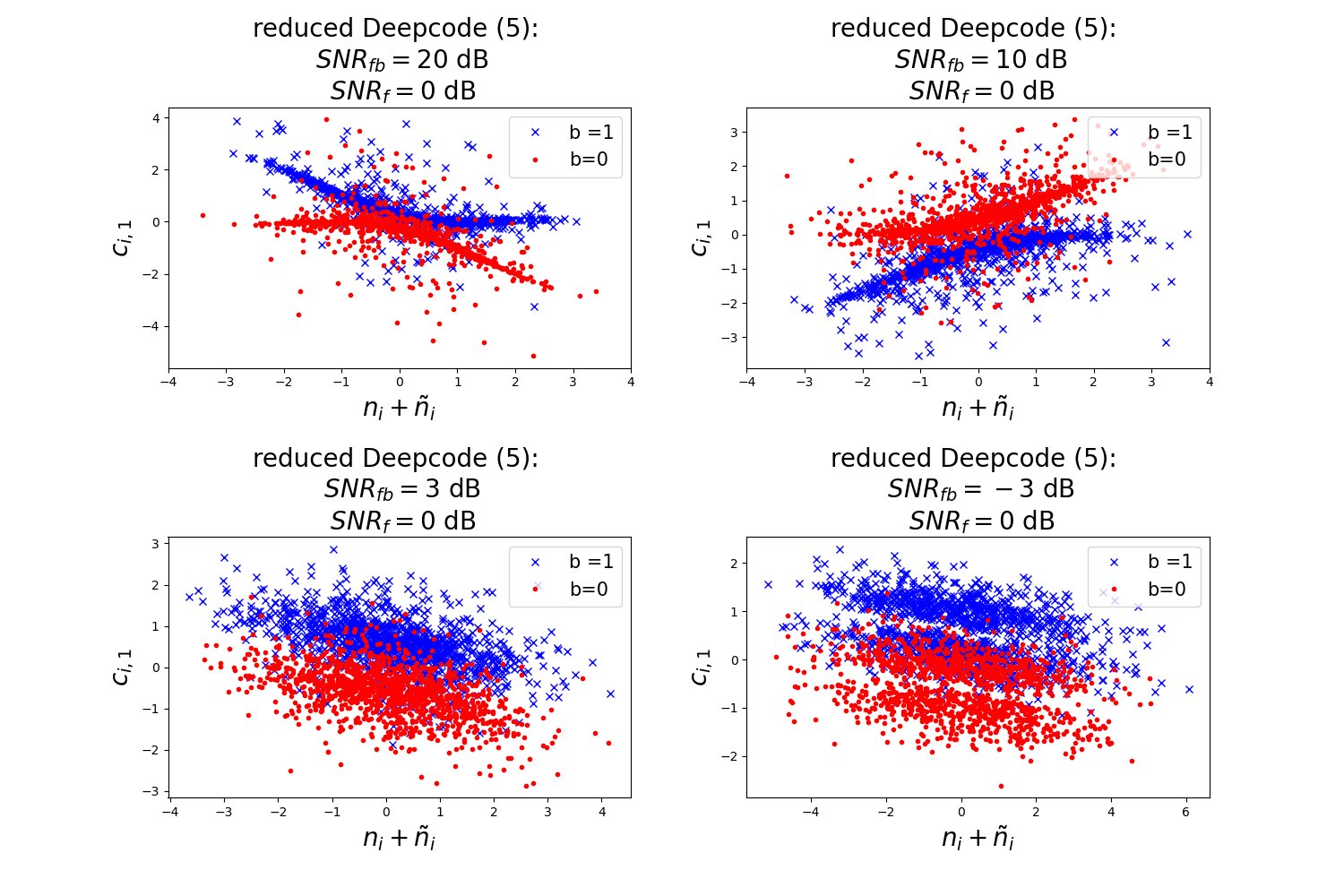}
    \caption{Reduced Deepcode (5): parity bit $c_{i,1}$ vs. sum of forward and feedback noises $n_i + \tilde{n}_i$ in the first phase}
    \label{fig:noisy parity}
\end{figure}

\begin{figure}[htbp]
    \centering
    \includegraphics[width=0.45\textwidth]{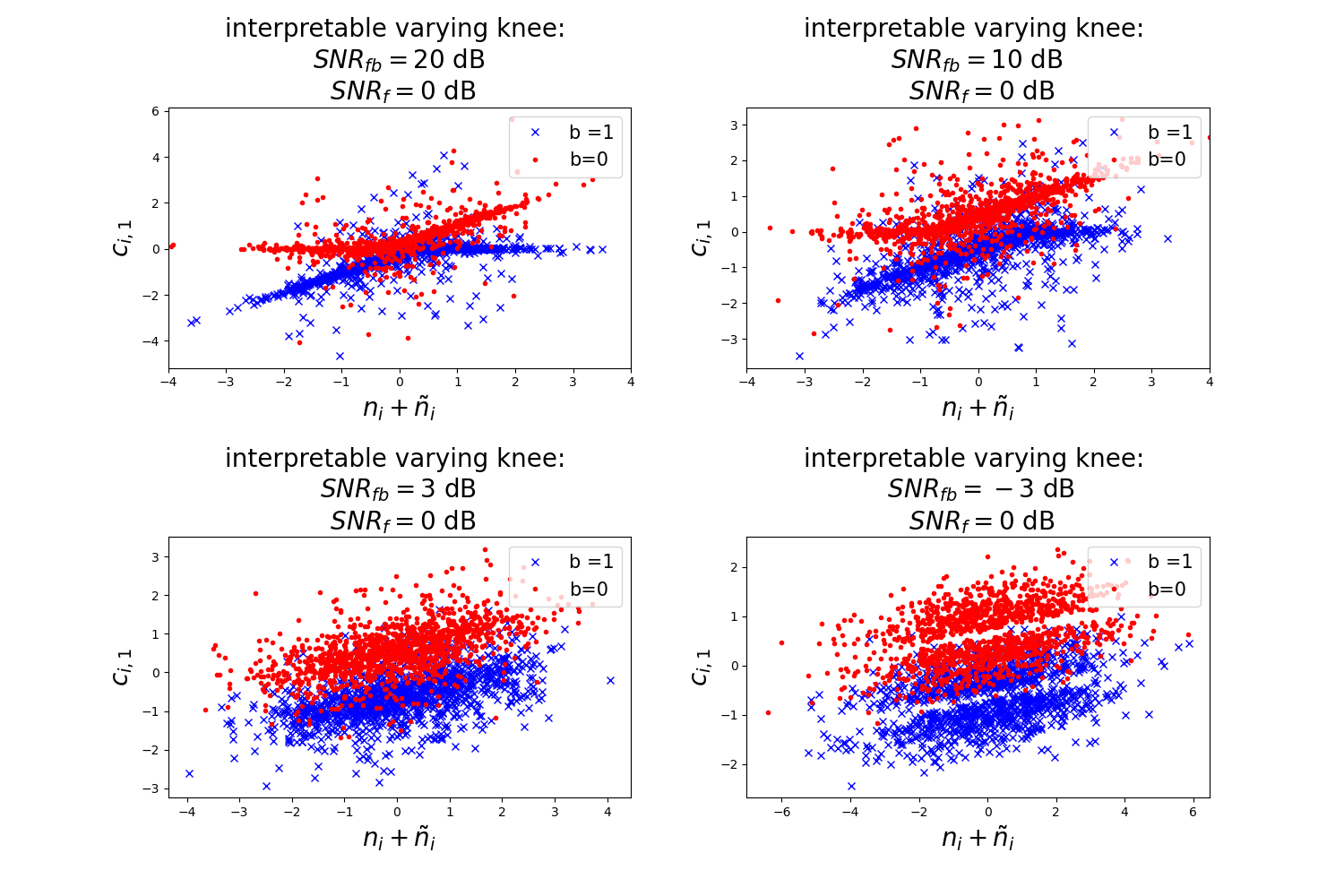}
    \caption{Interpretable varying knee: parity bit $c_{i,1}$ vs. sum of forward and feedback noises  $n_i + \tilde{n}_i$ in the first phase}
    \label{fig:noisy parity inter}
\end{figure}

When it comes to noisy feedback, Figs. \ref{fig:noisy parity} and \ref{fig:noisy parity inter} depict the changing pattern of codewords with decreasing feedback SNRs. Each scatter plot displays 2500 sample points: 50 samples for each $i$ where $i \in \{1, \ldots, 50\}$. By varying knee points, our interpretable model consistently aligns with the behavior of the reduced Deepcode ($5$). As the feedback SNR decreases, the knee points shift, and there are more outliers. This suggests that the model adapts to changing channel conditions by transmitting information about phase 1 noises across a wider range.

\section{Equivalent Codes}\label{apx: equivalent}


In this section, we discuss all the equivalent codes of our interpretable model. {In short, there is quite of bit of symmetry in  these codes, and the learned coefficients can be either positive or negative, as long as the combination of parity bits cancels out the noises.}

First, there are two options for selecting positive and negative signs:

$\bullet$ Type 1: as discussed previously, the parity bits may be formed as
\begin{align}
c_{i, 1} &= e_1n_{i}\mathbb{I}(-(2b_i-1)n_{i})  - e_2h_{i,4} - e_2 h_{i,5} \\
c_{i, 2} &= - e_1n_{i}\mathbb{I}(-(2b_i-1)n_{i}) - e_2h_{i,4} - e_2 h_{i,5}
\end{align}
In this case, we estimate $n_i$ by subtracting the current parity bits $c_{i, 1} \textcolor{red}{-} c_{i, 2}$ which only contain information about message bits and phase 1 noises. Error correction is then performed by adding the future parity bits $c_{i+1, 1} \textcolor{red}{+} c_{i+1, 2}$ which only contain $h_{i+1,4}$ and $h_{i+1,5}$. Their outliers will be used to assist decoding.

$\bullet$ Type 2: the parity bits can be expressed equivalently in another form.
\begin{align}
c_{i, 1} &= e_1n_{i}\mathbb{I}(-(2b_i-1)n_{i})  - e_2h_{i,4} - e_2 h_{i,5} \\
c_{i, 2} &= e_1n_{i}\mathbb{I}(-(2b_i-1)n_{i}) + e_2h_{i,4} + e_2 h_{i,5}
\end{align}
In contrast to before, we now estimate $n_i$ by adding the current parity bits $c_{i, 1} \textcolor{red}{+} c_{i, 2}$. The error correction term $h_{i+1,4}$ and $h_{i+1,5}$ are obtained by subtracting the future parity bits $c_{i+1, 1} \textcolor{red}{-} c_{i+1, 2}$.

Secondly, both $h_{i,4}$ and $h_{i,5}$ can take values of either $+1$ or $-1$ (4 possible combinations), with the sign of the ``\textcolor{cyan}{blue}'' portion changing correspondingly, enabling the values to cancel each other out when no outlier is present. Take Type 1 as an example, the values of $h_{i,4}$ and $h_{i,5}$ can be:

$\bullet$ Choice 1 $h_{i,4} = +1$ and $h_{i,5} = - 1$: $- e_2h_{i,4} - e_2 h_{i,5}$

$\bullet$ Choice 2 $h_{i,4} = -1$ and $h_{i,5} = - 1$: $+ e_2h_{i,4} - e_2 h_{i,5}$

$\bullet$ Choice 3 $h_{i,4} = +1$ and $h_{i,5} = + 1$: $- e_2h_{i,4} + e_2 h_{i,5}$

$\bullet$ Choice 4 $h_{i,4} = -1$ and $h_{i,5} = + 1$: $+ e_2h_{i,4} + e_2 h_{i,5}$

The BER performance  for each of these trained symmetric combinations is shown in Table. \ref{symmetry}. They all achieve the same level of BER performance, indicating that Deepcode exhibits symmetry and all codes are more or less equivalent. 

\begin{table}[htbp]
\centering
\begin{tabular}{|l|l|l|l|}
\hline
sign & $h_{i,4}$ &$h_{i,5}$ & BER  \\ 
\hline
Type 1     & $+1$   & $-1$          & $7.5951e-05$          \\ 
\hline
Type 1    & $-1$      & $-1$         &$7.8670e-05$            \\ \hline
Type 1     & $+1$     & $+1$          & $7.7344e-05$          \\ 
\hline
Type 1      & $-1$    & $+1$         &$7.8297e-05$          \\ 
\hline
Type 2      & $+1$    & $-1$          & $7.8231e-05 $          \\ 
\hline
Type 2     & $-1$     & $-1$         &$7.7326e-05$            \\ \hline
Type 2    & $+1$     & $+1$          & $7.9720e-05
$          \\ 
\hline
Type 2    & $-1$     & $+1$         & $7.7925e-05  $ \\
\hline
\end{tabular}
\caption{\label{symmetry}Equivalent codes at $\text{SNR}_f$ $0$ (noiseless feedback)}
\end{table}

\section{Noisy Feedback} \label{apx: noisy feedback}
In this section, we discuss the outlier analysis and BER performance in the case of noisy feedback.

\begin{figure}[ht]
    \centering
    \includegraphics[width=0.45\textwidth]{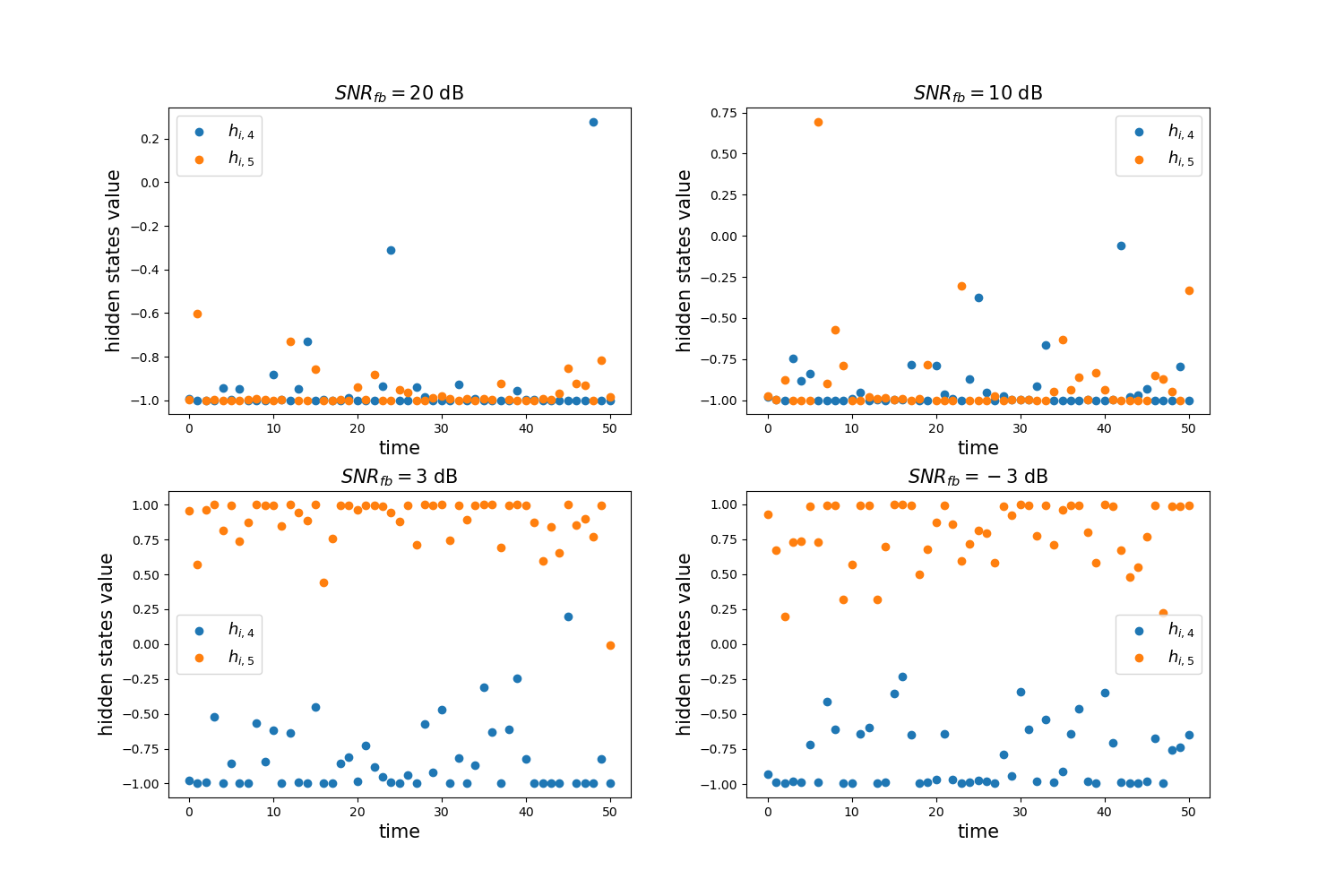}
    \caption{Outliers for noisy feedback ($\text{SNR}_{f} =0$)}
    \label{fig:outlier_noisy}
\end{figure}

Fig. \ref{fig:outlier_noisy} shows the values of $h_{i,4}$ and $h_{i,5}$ across various feedback SNRs. As expected, when the feedback SNR is lower, there will be more outliers available for error correction.

\begin{figure}[ht]
    \centering
    \includegraphics[width=0.45\textwidth]{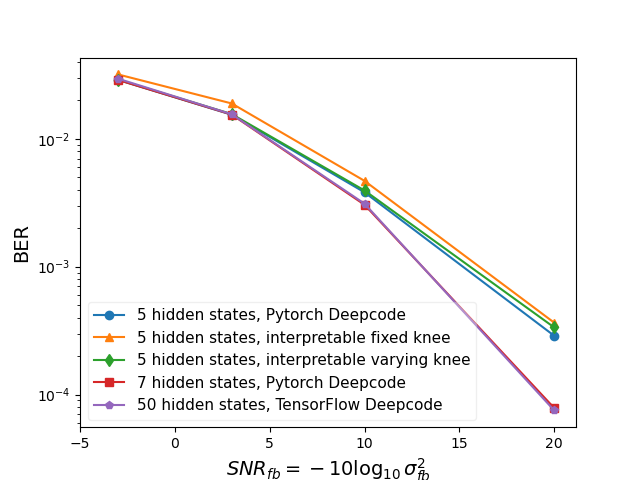}
    \caption{BER performance of models (noisy feedback, $\text{SNR}_f = 0$)}
    \label{fig:noisy ber}
\end{figure}

Fig. \ref{fig:noisy ber} shows the BER performance of the noisy feedback. 
In the Deepcode model, at high $\text{SNR}_{fb}$, the $5$ hidden states Deepcode performs worse than the $7$ and $50$ hidden states models. However, at low $\text{SNR}_{fb}$, despite having more hidden states for improved error correction, the $7$ and $50$ hidden states models do not outperform the $5$ hidden states model in terms of BER.
 We believe this is because Deepcode is sensitive to lower feedback SNRs. More hidden states deal with the forward noises more effectively than the feedback noises.


The generalized interpretable model with varying knee points performs comparably to the $5$ hidden states Deepcode, slightly outperforming the model with fixed knee points.  

\end{document}